\documentclass[twocolumn]{aastex62}

\submitjournal{PASP}

\shorttitle{A Comprehensive Astrometric Calibration of WFPC2}
\shortauthors{Casetti-Dinescu et al.}

\begin{document}

\title{A Comprehensive Astrometric Calibration of HST's WFPC2  : I. Distortion Mapping}

\correspondingauthor{Dana I. Casetti-Dinescu}
\email{casettid1@southernct.edu}

\author[0000-0001-9737-4954]{Dana I. Casetti-Dinescu}
\affiliation{Department of Physics, Southern Connecticut State University, 501 Crescent Street, New Haven, CT 06515}

\author{Terrence M. Girard}
\affiliation{Department of Physics, Southern Connecticut State University, 501 Crescent Street, New Haven, CT 06515}
\affiliation{Department of Astronomy, Yale University, Steinbach Hall, P.O. Box 208101, New Haven, CT 06520-8101}

\author{Vera Kozhurina-Platais}
\affiliation{Space Telescope Science Institute, Baltimore, MD 21218}

\author{Imants Platais}
\affiliation{Department of Physics and Astronomy, The Johns Hopkins University, Baltimore, MD 21218}

\author{Jay Anderson}
\affiliation{Space Telescope Science Institute, Baltimore, MD 21218}

\author{Elliott P. Horch}
\affiliation{Department of Physics, Southern Connecticut State University, 501 Crescent Street, New Haven, CT 06515}







\begin{abstract}

  Wide field planetary camera 2 (WFPC2) exposures are already some 20 years older than {\it Gaia} epoch observations,
  or future JWST observations. As such, they offer an unprecedented time baseline for high-precision proper-motion studies, provided
  the full astrometric potential of these exposures is reached. We have started such
  a project with the work presented here being its first step. We explore geometric distortions beyond the well-known ones published
  in the early 2000s. This task is accomplished by using the entire database of WFPC2 exposures in filters F555W, F606W and F814W
  and three standard astrometric catalogs: {\it Gaia} EDR3, 47 Tuc and $\omega$Cen. The latter two were constructed using $HST$ 
  observations made with cameras other than WFPC2. We explore a suite of centering algorithms, and various distortion maps in order to 
  understand and quantify their performance.

  We find no high-frequency systematics beyond the 34th-row correction, down to a resolution of 10 pixels. Low-frequency systematics 
  starting at a resolution of 50-pixels
  are present at a level of 30-50 millipix (1.4-2.3 mas) for the PC and 20-30 millipix (2-3 mas) for the WF chips. We characterize these 
  low-frequency systematics by providing correction maps and updated cubic-distortion coefficients for each filter.
\end{abstract}

\keywords{Astrometry: space astrometry --- Stellar kinematics: proper motions}


\section{Introduction} \label{sec:intro}
We live in an era where precision astrometry is revolutionizing our understanding of the local universe.
Much of this recent leap forward is due to two space-based platforms, {\it Gaia} and the Hubble Space Telescope ({\it HST}).
While {\it Gaia} was designed from the start as an astrometric mission, {\it HST} had a broader
scope, with its imaging capabilities, in particular, playing a major role in numerous scientific discoveries.
After a meticulous calibration effort geared toward astrometry, {\it HST} has also proved to be a high-precision astrometric 
instrument, poised to compete with and complement {\it Gaia}.
For instance, {\it HST} has provided accurate
measurements of distant Milky-Way satellites and globular clusters
\citep[e.g.,][]{sohn17,sohn18,kall13},
internal motion and rotation in globular clusters
\citep[e.g.,][]{wat15,bell17},
as well as parallaxes of various
objects of interest
\citep[see review by][]{ben17,riess16}.

{\it HST} has three advantages over {\it Gaia}: better resolution in crowded stellar fields, depth, and a longer time baseline
for motion studies. {\it Gaia}, of course has the advantage of full-sky coverage and unprecedented astrometric precision at 
magnitudes that overlap well with a large portion of {\it HST} exposures.
A comparison of {\it Gaia} DR2 \citep{g18a} proper-motion measurements of distant Milky-Way satellites with measurements made by
various groups using {\it HST} data is presented in \citet{g18b}.
Figure 15 of \citet{g18b} makes it abundantly clear that the best agreement between the two platforms is obtained when
{\it HST}
measurements 1) have long time baselines ($\sim 10$ years) and 2) are made with thoroughly-calibrated instruments such as the 
ACS/WFC
and the WFC3/UVIS.

Here, we will combine the attributes of {\it HST} and {\it Gaia} in order to better calibrate, astrometrically, older {\it HST} 
observations and thus make these available for a variety of proper-motion studies.
Specifically, we calibrate the Wide Field Planetary Camera 2 (WFPC2), making use of all appropriate observations taken
since it was installed (1993) up to its decommissioning (2009).
The Mikulski Archive for Space Telescopes (MAST) includes WFPC2 observations of $\sim 100$ Milky-Way globular clusters and many 
regions of interest in the Magellanic Clouds
and in the Galactic bulge, to name a few, thus offering a rich database of early-epoch astrometry for proper-motions studies.
Furthermore, early-epoch, accurately calibrated WFPC2 positions can be combined not only with existing modern {\it HST} 
observations, but
with new observations taken from the ground with various high-resolution imaging techniques,
\citep[e.g., the Gemini multiconjugate adaptive optics system,][]{mass16,patti19}
as well as upcoming observations with {\it JWST}.

\section{Data Sets} \label{sec:data}
The calibration scheme we develop heavily depends on the number of accumulated astrometric solutions 
and thus implicitly, observations.
Therefore, we will use observations taken in filters F555W, F606W and F814W that have the largest archived datasets.
An inventory of the exposures is presented in Table \ref{tab:obs_census}.
We have used the MAST to obtain the corresponding processed fits files.
These data were kindly shipped to us on a hard drive
upon special request to MAST personnel. 
Comparing the MAST list of WFPC2 observations and the archived files, we find that a handful of exposures are missing from 
the archives: three in F555W and one in F814W.

\begin{figure}[!h]
\includegraphics[angle=0,scale=0.6]{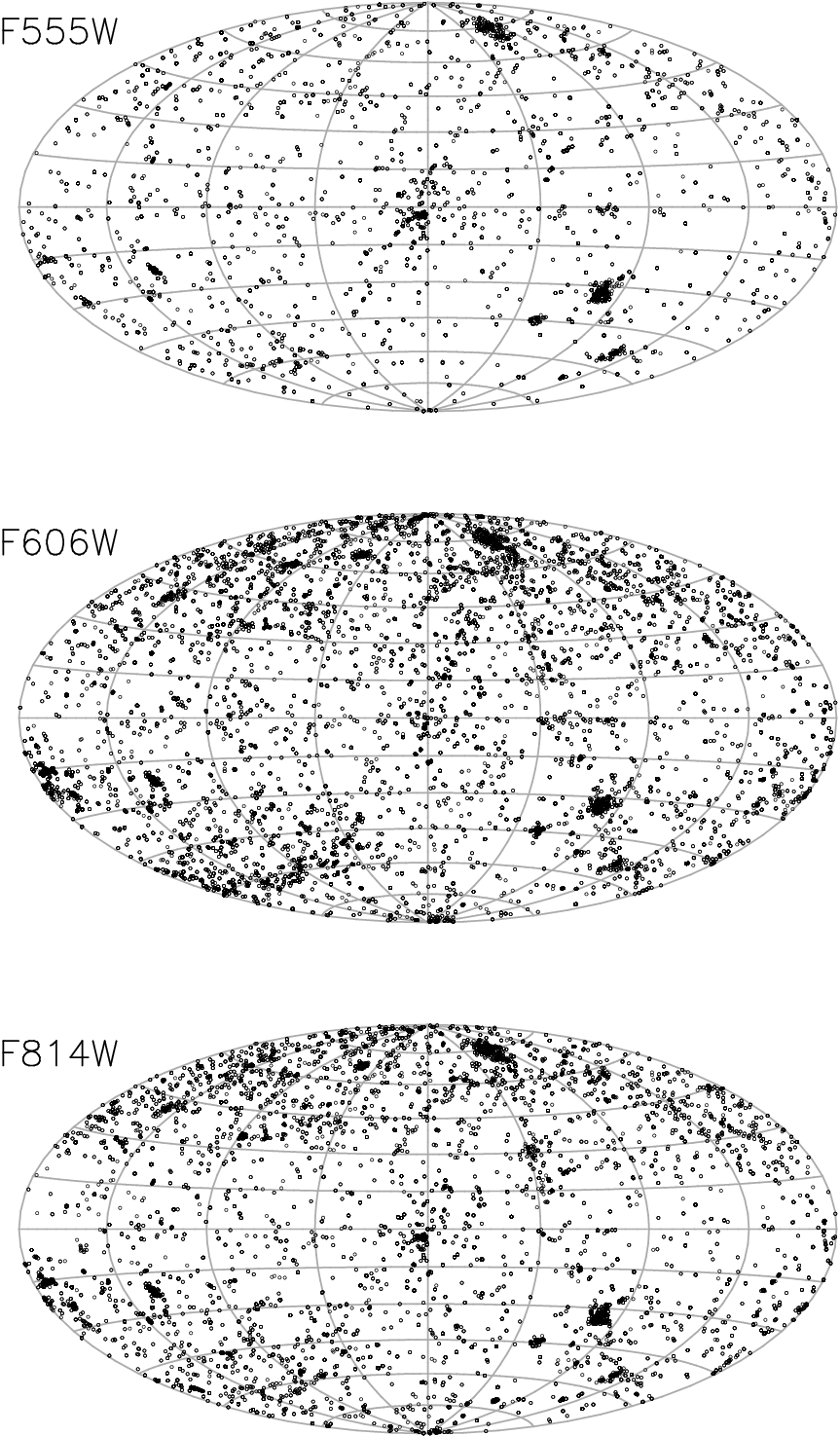}
\caption{Galactic-coordinates distribution of observations in each filter. For reference,
  LMC-related observations cluster around $b= -33\arcdeg$, and SMC-related ones around $b= -45\arcdeg$.
\label{fig:gal_distrib}}
\end{figure}

We also choose not to use any binned exposures, thus eliminating 9 more exposures in F555W and 18 in F814W. The total number of 
exposures listed in column 2 of Tab. \ref{tab:obs_census} thus represents the number of usable exposures that were considered in 
this study; the following columns indicate the number of exposures per each WFPC2 chip.
The distribution of the observations in Galactic coordinates is shown in Figure \ref{fig:gal_distrib} for each filter. 
Distributions are highly correlated, especially in filters F555W and F814W, indicating the photometric scientific purposes of the 
vast majority of these observations.

\begin{table}
\caption{List of WFPC2 Exposures }
\begin{tabular}{rrrrrr}
\hline
\multicolumn{1}{c}{Filter} & \multicolumn{1}{c}{Total} & \multicolumn{1}{c}{PC} & \multicolumn{1}{c}{WF2}  & 
\multicolumn{1}{c}{WF3} & \multicolumn{1}{c}{WF4} \\
\hline
F555W & 14828 & 14312 & 12914 & 13008 & 13031 \\
F606W & 28257 & 28191 & 28186 & 28249 & 28185 \\
F814W & 26466 & 26309 & 25507 & 25555 & 25499 \\
\hline
\end{tabular}
\label{tab:obs_census}
\end{table}


\section{Aims and Overall Strategy} \label{sec:aims}
The primary purpose of this study is to identify sources of systematic errors in WFPC2 astrometry and model them as best possible. 
To this end, we are not particularly concerned with detection completeness and/or photometric accuracy.  Achieving these 
requires a specific treatment for each set of observations and this was likely done when the original scientific analysis of a given 
data set was performed. We focus on building a standardized astrometric procedure that works reasonably well in most circumstances: 
crowded 
versus sparse fields, shallow versus deep exposures, etc. Naturally, sets of exposures that lack sufficient stars to perform 
transformations, one into another or into a standard catalog such as {\it Gaia}, will be of no use to our purpose. Such exposures 
include high Galactic latitude regions, very shallow exposures, and exposures focused on bright, extended objects such as 
solar-system planets and such.

Systematic, geometric errors are determined by stacking residuals from multiple ``plate'' solutions, i.e., transformations between
detector pixel coordinates and standard-catalog celestial coordinates.
The celestial coordinates are first transformed into standard coordinates, $(\xi, \eta)$, via gnomonic projection about a suitable
tangent point for each exposure, and then rotated to align with the $(x, y)$ pixel system of each chip based on the WCS information
of the WFPC2 fits files.
Thus, residuals from the transformations will be in the detector coordinate system.
We adopt a general cubic polynomial transformation between rotated standard coordinates and detector positions: \\ \\
$\xi_r  = a_0 + a_1 x + a_2 y + a_3 x^2 + a_4 x y + a_5 y^2 + a_6 x^3 + a_7 x^2 y + a_8 x y^2 + a_9 y^3  $ $~~~~(1a)$ \\ \\
$\eta_r = b_0 + b_1 y + b_2 x + b_3 y^2 + b_4 y x + b_5 x^2 + b_6 y^3 + b_7 y^2 x + b_8 y x^2 + b_9 x^3  $ $~~~~(1b)$ \\ \\
where $\xi_r, \eta_r$ are the rotated standard coordinates of a reference star and $x, y$ are the star's position in the detector
relative to a fiducial position, in pixels, taken to be (425, 425).
The twenty coefficients, $a_n$ and $b_n$, are determined by an iterative least-squares procedure with outlier culling.
Residuals are the differences between the standard catalog positions and calculated positions for all reference stars.

We investigate and model systematic errors, grouped by amplitude and by scale length over which they operate. For instance, the 
well-documented known systematics of WFPC2 are the 34th-row correction \citep[][hereafter AK99]{ak99}, and the geometric 
optical distortion \citep[][hereafter AK03]{ak03}. 
The former operates with an amplitude of 15 millipix (mpix, thereafter) over some 34 pixels of the 
detector and is due to a manufacturing defect in all four chips; one of every 34 rows is narrower than the rest. The latter 
operates with an amplitude of $\sim 4$ pixels over some 400 pixels (i.e., from center to edge of a chip). Systematics that operate 
on scales smaller than or comparable to the 34th-row correction we will call high-frequency, while those larger than the 34th row, 
we will refer to as low-frequency. Our primary goal is to explore whether (geometric) systematics of high and low frequency exist 
beyond what was found by AK99 and AK03. We will also explore the variation of these systematics with time. The charge transfer 
efficiency (CTE) \citep{dol09} effect on positions will be explored in a following study.

Each WFPC2 chip is modeled separately as an individual unit. We use internal calibrations, meaning relative coordinate 
transformations between overlapping sets of exposures, as well as external calibrations, meaning coordinate transformations between 
WFPC2 exposures and a standard catalog. We make use of three external standard catalogs: {\it Gaia} EDR3 \citep{g20}, one near
47 Tuc and based on ACS/WFC observations in an off-center field of the cluster \citep{koz15b,koz18}, later placed on the 
{\it Gaia} DR2 system \citep{g18a}, and a third catalog of $\omega$ Cen based on WFC3/UVIS observations in the center of the 
cluster \citep{koz15a}.
Details are provided in the following section on standard catalogs.

In the case of internal calibrations, the standard coordinates on the left-hand side of Equations 1a and 1b are replaced by the $(x, y)$
detector coordinates from a chosen reference frame, be it a single exposure or the average of multiple exposures transformed
onto a common reference system.  
Throughout, we shall specify whether residuals are derived from external or internal calibrations.
In general, internal calibrations are used for tests of precision and consistency checks while our final distortion coefficients
and maps are based solely on calibrations with external catalogs.

With the exception of some specific tests, all WFPC2 coordinates are first pre-corrected for the two known systematics: the 
34th-row (AK99) and the nominal 3-rd order optical distortion (AK03, also WPFC2 instrument handbook v10.0). The AK03 distortion 
correction makes use of a single set of cubic coefficients for all 
three filters, namely those in Table 5.6 of the WFPC2 instrument handbook. We will call this the {\it nominal} distortion. 
Distortion for filters F814W and F300W was explored by \citet{koz03} as a difference between the nominal distortion and the 
residuals provided by the new filters, much in the same way as we do here, but using far fewer data sets and no standard external 
catalog.

Here we determine distortion maps as post corrections to the nominal distortion of AK03, for each filter and two epoch ranges, 
using the three external standard catalogs mentioned above.

\section{Standard Catalogs} \label{sec:std_catalogs}
Basic properties of the external standard catalogs are listed in Table \ref{tab:std_cat}.
Our procedures were originally developed using {\it Gaia} DR2. 
With the recent release of {\it Gaia} EDR3 \citep{g20}, we were able to reprocess the entire data set quickly
and our results presented here are based on EDR3, although, occasionally we may refer to DR2 for some tests.
From {\it Gaia} EDR3 we make use of
all stars that have a proper-motion measurement.
Since each WFPC2 chip encompasses a rather small
sky area, we need every single {\it Gaia} star that has a reasonably well-determined
position, translatable to the epoch of the WFPC2 observation. 
The errors in the astrometric transformations of WFPC2 star positions into {\it Gaia} are 
dominated by the {\it Gaia} EDR3 proper-motion errors which propagate into positional errors when backdated 
to the epoch of the WFPC2 observations.  Accumulating residuals from such transformations over many pointings on the sky 
(see Fig. \ref{fig:gal_distrib}) ensures that any spatially-dependent systematics in {\it Gaia} 
\citep[e.g,][]{g20} --- much smaller in EDR3 than in DR2 --- will not impact are results.  Instead it is 
the random {\it Gaia} position errors at the WFPC2 observation epoch that dominate the error budget.
For instance, {\it Gaia} stars participating in WFPC2 solutions are typically faint, peaking at $G \sim 19 $, and thus have an 
approximate proper-motion error of $\sim 0.3$ mas~yr$^{-1}$. 
Propagating back twenty years (for early WFPC2 observations), this corresponds 
to 6-mas positional error, or 130 mpix for the PC and 60 mpix for the WF chips. 
Star-image centering errors are typically below these values for the WFPC2. 
Nevertheless, averaging hundreds of thousands of such residuals allows us to characterize systematic WFPC2 distortion
beyond that of the applied, nominal distortion correction.

The 47 Tuc catalog is the most precise and accurate among the three standard catalogs. It is based on ACS/WFC F606W observations 
\citep{koz15b} of an off-center field of the cluster.
The off-center pointing allows good overlap with {\it Gaia} measurements, despite its crowdedness limitations. The original
47 Tuc catalog was improved using {\it Gaia} DR1 \citep{g16} by \citet{koz18}. Kozhurina-Platais (private communication) used {\it 
Gaia} DR2 to further improve the positions; it is this version of the catalog with which we started to work. Upon inspection of 
a transformation of the catalog positions into {\it Gaia} DR2 positions, we noticed that, bright, well-measured stars in {\it Gaia} 
still presented significant 3-rd order terms.
Therefore, we have used {\it Gaia} DR2 well-measured, bright stars to adjust for residual distortion in the positions of the 
catalog. We have also used well-measured stars ($\sim 5500$ stars with $G \le 20$)
to place the relative proper motions of the catalog onto the absolute system of {\it Gaia} DR2. Finally,
for convenience in our exploration of systematics within magnitude ranges, we have adjusted the catalog's F606W instrumental 
magnitudes to the {\it Gaia} $G$ magnitude system. This was done by adding a constant to the 47 Tuc instrumental magnitudes.  
A plot of magnitude differences as a function of {\it Gaia} $(G_B-G_R)$ color showed no trends.  
The rms of the differences was 0.03 mag.
Proper-motion uncertainties in the 47 Tuc standard catalog have an upper limit of 0.50 mas~yr$^{-1}$, 
with an average of 0.12 mas~yr$^{-1}$.

The $\omega$ Cen catalog \citep{koz15a} is based on WF3/UVIS F606W observations in the center of the cluster. As such, a 
{\it Gaia}-based astrometric adjustment is not possible, as there are very few {\it Gaia} stars in this very crowded region, 
and those present are rather poorly measured
\citep[see also][]{koz18}. From this original catalog, we keep objects having at least 5 observations and with
proper-motion uncertainties $\le$ 10.0 mas~yr$^{-1}$.
To place the instrumental magnitudes on a standard system we do a match with {\it Gaia} DR2. Of the 
263 matches, we use 90 suitable stars (neither too faint nor too bright) to determine the magnitude offset (with an rms of 0.03 
mag). After applying the offset, we find that the faint limit of the catalog is close to the overall {\it Gaia} limit (see Tab. 
\ref{tab:std_cat}). Proper motions are relative, however this will not affect our solutions, since the vast majority of stars that 
will be used in WFPC2 transformations are cluster stars (and thus share a common systemic motion).
Mean proper-motion uncertainties are $\sim 0.14$ to 0.20 mas~yr$^{-1}$ for stars with $G > 16$. Brighter stars,
$G \sim 14 - 16$, have larger proper-motion uncertainties, with a mean of $\sim 1.0$ to 2.0  mas~yr$^{-1}$. 
The largest uncertainties are in $\mu_{\delta}$, in both magnitude ranges.

The advantage of the two cluster-based standard catalogs is that they are at epochs closer to the WFPC2 observations than is {\it 
Gaia} EDR3; the 47-Tuc catalog is substantially deeper and the most precise. These cluster-based standard catalogs will serve as
useful checks on the {\it Gaia} EDR3 results. In Tab. \ref{tab:std_cat} we also list the range of the number
of WFPC2 exposures used to perform suitable solutions into these catalogs,
as well as the range of the number of reference stars per chip used in such solutions.
The ranges reflect variations between filters and chips (see also Section \ref{subsec:low_freq}).

\begin{table*}
\caption{Standard Catalogs Properties }
\begin{tabular}{lllrlcc}
\hline
\multicolumn{1}{c}{Catalog} & \multicolumn{1}{c}{Instr.} &  \multicolumn{1}{c}{System} & \multicolumn{1}{c}{$<Epoch>$} & 
\multicolumn{1}{c}{$G$-limit} & \multicolumn{1}{c}{\# WFPC2 exposures} &  \multicolumn{1}{c}{\# ref stars/chip} \\
\hline
{\it Gaia} EDR3 & {\it Gaia} & absolute & 2016.0 & $ \sim 21.0$ & 2600 - 5200 & 20 - 600 \\
47 Tuc - off cen. & ACS & absolute & 2007.8 & $ \sim 25.3$ & 32 - 132 & 50 - 1600 \\
$\omega$ Cen - center & WF3 & relative  & 2011.2 & $ \sim 21.2$ & 18 - 26 & 400 - 4700 \\
\hline
\end{tabular}
\label{tab:std_cat}
\end{table*}

\newpage
\section{Centering Algorithms} \label{sec:main_centering}
In this section various centering algorithms are considered and their performance on WFPC2 stellar images evaluated.
We feel it is important to convey what we have learned in this regard, although admittedly the details may not be of
interest to all readers.  For this reason we state that the two centering methods adopted for determining our final
distortion maps and corrections to the cubic-distortion terms are 2-D Gaussian centering and the effective PSF (ePSF) algorithm,
as detailed below.
Discussion of the derivation of distortion corrections resumes with Section \ref{sec:dist_maps}.

The best choice of algorithm for centering WFPC2 stellar images is not obvious, given our intended purpose.
Our goal is to minimize the noise in the average residuals generated by transformations into the standard catalogs.
When using {\it Gaia} EDR3 as the standard catalog, propagation of the EDR3 proper-motion errors to the WFPC2 epoch will be the
dominant source of scatter in the residuals, provided the WFPC2 centers are even moderately well-determined.
In which case it is important to choose a centering algorithm that performs over a wide range in magnitude (even into
the realm of saturated images) in order to maximize the number of {\it Gaia} stars that can be utilized as reference stars.
Conversely, when employing the 47 Tuc and $\omega$ Cen standard catalogs, centering precision of the WFPC2 stellar images is
paramount, given the higher precision of these external catalogs and the relatively few WFPC2 exposures available in
these areas. 
Finally, there is the practical matter of computational speed to consider.

In general, the process of determining precise centers of stellar images consists of three steps:
pre-processing the image data to properly calibrate for detector response,
detecting stellar sources as groups of pixels with signal above the background noise,
and then refining the positions of each source by fitting the nearby pixels with some model function.
The MAST provides calibrated $_{-}$c0m.fits files for every WFPC2 exposure, i.e.,
images that are bias-subtracted, dark-subtracted, and flat-fielded.
Together with each target image, a bad-pixel mask is also provided ($_{-}$c0m.fits files).
We split these multi-extension FITS files and treat each chip as a separate entity.

For most 
of the centering methods that we explore the source-detection step is performed using
the code Source Extractor (SExtractor, version 2.19.5) developed by \citet{bert96,bert10}.
Based on an image and its corresponding bad-pixel mask, SExtractor detects objects and provides preliminary centers, 
instrumental magnitudes, and other image parameters. 
For WFPC2 images, regions along two edges of each chip are spoiled by vignetting from the four-faceted pyramid mirror and these
are among the marked pixels in the bad-pixel masks.
Similar limits are are also given in the WFPC2 instrument handbook.
However, we found the astrometry degrades over a somewhat wider region along these edges than is suggested by the masks
and the WFPC2 handbook and, thus, we adopt a more restricted usable area.
Specifically, for all chips we only accept sources located between 75 and 795 pixels, in both axes. 
After preliminary tests using a range of SExtractor parameter values, we decided on a small size for the local
background (BACK$_{-}$SIZE = 16 pixels) in order to detect as many objects as possible in crowded areas with elevated 
backgrounds, such as cluster cores. The input FWHM parameter chosen varies by chip and filter. For the PC we have used $1.6, 1.6, 
1.8$ pix for filters F555W, F606W and F814W, respectively.  For the WF we have used
$1.2, 1.3, 1.4$ pix for filters F555W, F606W and F814W, respectively. These values were determined from the peaks of the 
distributions of the FWHM for each chip and filter.

A number of image fitting methods were explored; each is described in the following subsections.

\subsection{2D Gaussian Centering} \label{subsec:2DG}
Two-dimensional, elliptical Gaussian functions (2DG) are fit to the intensity profile of each object. We use the Yale centering 
routines upgraded for CCD data from the original code \citep{lva83}, with centers from SExtractor serving as initial values for the 
non-linear fitting algorithm. Compared to the other methods tested, this algorithm is fast, and performs well for bright, 
partially saturated stars, enabling the largest overlap with {\it Gaia} EDR3 stars.
We note that the analytical form of this fitting function allows it to vary its width and shape across the chip.

\subsection{ePSF Centering} \label{subsec:epsf}
The ePSF code described in \citet{ak00} utilizes
an {\it effective} PSF built empirically from observations and has become somewhat of an {\it HST} standard.
This code (hst1pass.2019.10.11) includes its own detection algorithm and also provides 
object centers, instrumental magnitudes and a quality-of-fit parameter (q).  
The mode in which we run the program employs an existing library of ePSFs specifically constructed for the WFPC2; these were
provided along with the software.
For each chip, a $3 \times 3$ x-y grid of ePSFs is contained in the library.
WFPC2 ePSF libraries exist for only two of our filters: F555W and F814W. Thus, for F606W, we have used the F555W library ePSFs. 
As with the other centering methods, only sources in the area between 75 and 795 pixels, in both axes, are accepted.

The ePSF algorithm is fast and provides the best object centers for the WF as we will further detail. Because the ePSF algorithm
fits only the inner $ 5 \times 5 $ pixels of an object, the code tends to discard a large number of bright stars that are nearing
saturation but are otherwise measurable, astrometrically. 
Where the code really shines is in providing very high precision centers for fainter stars. 
As a consequence, there will be a limited magnitude-range
overlap with {\it Gaia} EDR3 stars, and this overlap is where errors in {\it Gaia} EDR3 increase rapidly. 
For reference, we use only objects with q between 0.0001 and 1.0.


\subsection{Dolphot Tiny Tim PSF Centering} \label{subsec:TT}
The DOLPHOT package \citep{dol00} version 2.0 was used to test the astrometric performance of image centers determined 
with Tiny Tim PSFs. These are optical-model PSFs based on the {\it HST} TinyTim simulator \citep{krist11} for each camera, chip and 
filter. Pre-processing included masking bad pixels, separating into the four chips and calculating the sky background, 
all done following the Dolphot manual recommendations. Consistent with previous schemes, we use 
only images detected within 75 and 795 pixels (both coordinates) of each chip.
The algorithm is geared toward optimizing source detection and photometry but is
prohibitively slow for our intended application.

\subsection{PSFEx} \label{subsec:psfex}
PSFEx (PSF Extractor) is a code developed by \citet{bert10} that constructs empirical models of a spatially varying PSF from 
images pre-processed with SExtractor. 
We tested version 3.17.1 of this code. To build the PSF library, we use the 47-Tuc core F555W data set of 636 160-sec exposures 
that were offset by no more than 1 WF pixel (PI Gilliland, PID 8267) taken in July 1999. 
Objects detected with Sextractor are candidates for PSF 
computation: we further limit these to a FWHM range between 0.8 and 3.0 pixels, a minimum SNR = 50 and a maximum ellipticity of 0.2.  
Each object 
thus selected, is cut out from the image with a square raster size of 5x5 pixels (the ``vignetting'' parameter in PSFEx) and then 
fed into the PSF-model builder. Our selection yielded $\sim 162,000$ PSF stars for the PC, and $\sim 223,000$ PSF stars for the 
WF chip. The PSF stars are well-distributed over the chip area. To account for spatial variation, we use a 5x5 $x-y$ grid of PSF 
models across each chip and linear interpolation.
Each PSF is built in super-resolution mode, utilizing a PSF size of 7x7 pixels. 
The parameter controlling the under/oversampling of the 
PSF is known as ``sample'' in the PSFEx code; we explored a range of values between 0.8 and 3.0 with a step of 0.2. 

Once the PSF 
model is built, SExtractor is run again to obtain new centers based on PSF-fitting for all detections in our target exposure,
which was a 1400-sec F555W exposure of 47 Tuc (PID 6114, epoch 1995.815) that overlaps with the 47 Tuc standard catalog. 
Astrometric solutions into this 
catalog, including up to third-order polynomial terms, were performed. 
The best solutions obtained were for PSFs using a sampling factor of 1.2 
for the PC and 2.0 for the WF. That is, these values resulted in the smallest standard errors of the transformation while
retaining a reasonable 
number of stars in the solution. The centers were found to have a small pixel-phase bias 
(see below, in Section \ref{subsec:pixelphase}). 
The algorithm is relatively cpu-intensive, which prevented our exploration of any possible variation of the PSF library with epoch
of observation.
This latter 
aspect also affects the ePSF algorithm which has a library based on a single-epoch data set.

\subsection{Fourier technique + 2D Gaussian} \label{subsec:fourier}
This procedure is somewhat different in that it applies additional pre-processing before the image detection and fitting steps.
It begins by creating an idealized diffraction-limited image of a point source using the known pixel scale, the 
parameters of the primary aperture with central obscuration of the secondary mirror, and the wavelength of observation. 
The resulting image is 
then used to deconvolve the WFPC2 exposure in the Fourier domain, resulting in a true estimate of the Fourier transform of the 
image. However, at higher spatial frequencies, the noise in the Fourier domain begins to dominate, and therefore a low-pass filter 
is applied. This filter is a Gaussian filter where the wings have been suppressed by multiplying by a Butterworth filter and 
thresholding.  This effectively sets extremely low but non-zero values to zero. 
After application of the filter, the resulting frame is then inverse-transformed to arrive at a Fourier-reconstructed image. 
Finally, this reconstructed image is processed 
as in Sec. \ref{subsec:2DG}. The Fourier technique has the advantage of suppressing the pixel-phase bias. However, faint objects 
are lost in the Fourier reconstruction (see Secs.
\ref{subsec:pixelphase}, \ref{subsec:trans_std_catalog}).

\section{Comparative assessment of the centering algorithms} \label{sec:comparative}
To illustrate the head-to-head centering performance of the various algorithms we select an appropriate WFPC2 exposure within
the area of the 47 Tuc standard catalog, which is the
deepest and most precise of the standard catalogs (Sec. \ref{sec:std_catalogs}).
We choose an F555W 1400-sec exposure taken at epoch 1995.815 (PID 6114), this wavelength passband being the most undersampled
of the three filters considered, i.e., the most demanding scenario.
Two aspects of the various centers are examined: the astrometic precision as measured by a 
transformation into the 47 Tuc standard catalog positions, and a diagnosis of the centering algorithm's pixel-phase bias.

Note that the test results from the transformation into an external standard catalog, while providing uniformity in the
comparison, will include errors of this external catalog. 
These errors are largely the proper-motion errors of the external standard catalog propagated into positions at the epoch of the WFPC2 exposure. 
In a future study, we will present a more in-depth analysis of the centering precision as also
determined from relative astrometry, i.e., from repeated, offset WFPC2 exposures.

\subsection{Astrometry from Transformation into a Standard Catalog}\label{subsec:trans_std_catalog}
Each set of centers for this test exposure are transformed into the epoch-adjusted, gnomonic projected and rotated 
celestial coordinates of the 47 Tuc standard catalog using Eq.~1 and the procedure described in Sec. \ref{sec:aims}.
We note that while 2DG, ePSF and Fourier positions were pre-corrected with the 34th-row correction and the nominal
distortion, the Dolphot/TinyTim and PSFEx positions were not. This does not affect the conclusions of this particular test since 
third-order terms are included in the transformations, thus absorbing residual distortion, while the 34-row issue will make
a negligible contribution to the overall scatter of the residuals, 
its amplitude being at most 15 mpix and only affecting a very small subset of images, those that happen to fall on such rows.
The residuals of the resulting transformations will be the quadrature sum of the centering errors 
of each algorithm on an WFPC2 exposure and the position errors of the external standard catalog.
The latter are dominated by the proper-motion errors of the 47 Tuc standard catalog propagated to the epoch of the WFPC2 
observation, i.e., some 12 years.
Considering an average proper-motion uncertainty of 0.12 mas/yr (Sec. \ref{sec:std_catalogs}), the position errors of the
standard catalog are $\sim 31$ mpix for the PC and $\sim 14$ mpix for the WF chips. 

\begin{figure*}[h]
\includegraphics[angle=-90,scale=0.67]{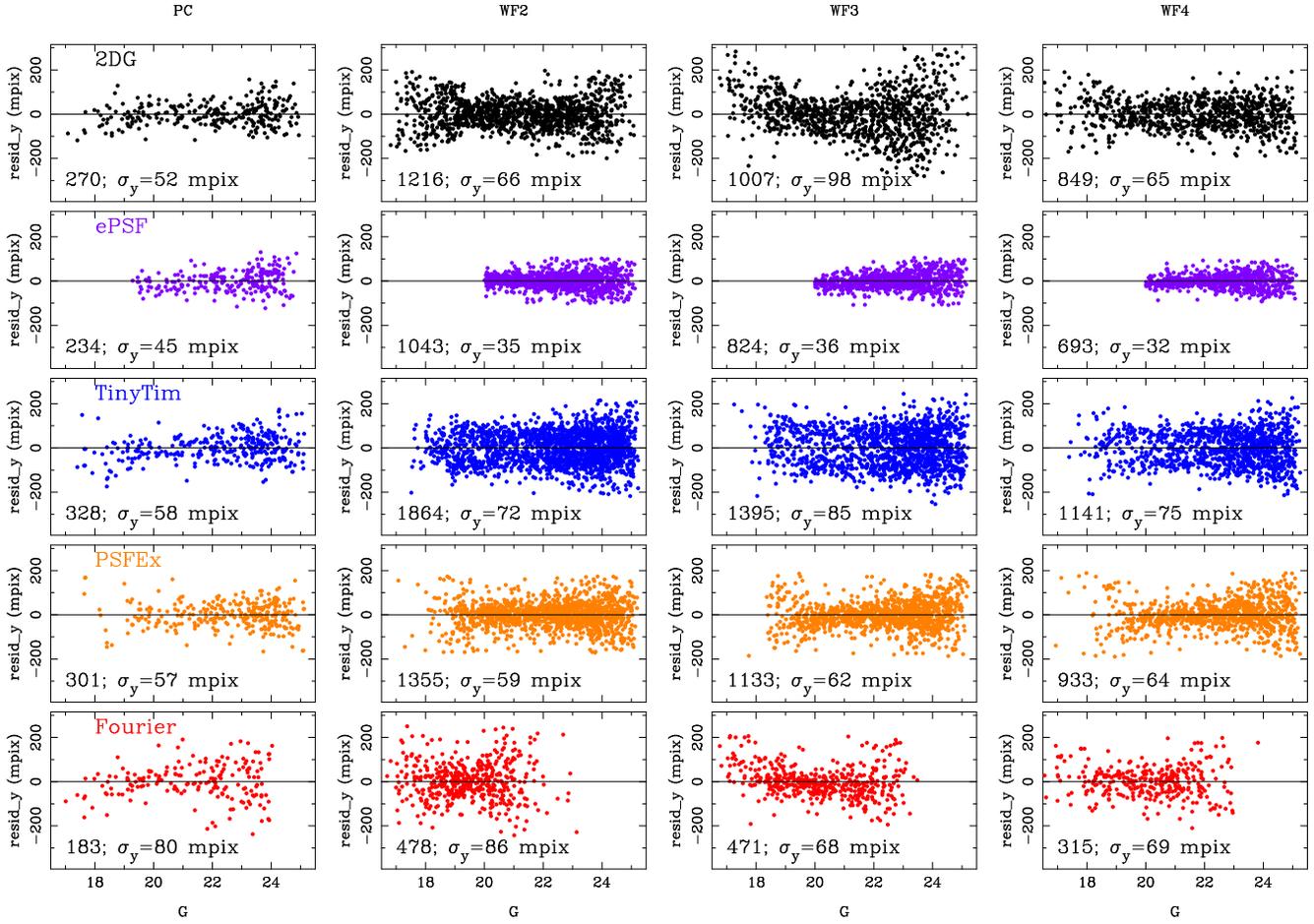}
\caption{Residuals from the transformation of star centers from an F555W, 1400-sec exposure (PID 6114) into
  the 47 Tuc standard catalog as a function of $G$ magnitude. Each panel represents a centering algorithm and chip as specified. 
  The number of stars participating in the solution and the standard error of the solution are also indicated for each case.
  The contribution of the external catalog position errors to the error budget is $\sim 31$ mpix for the PC and $14$ mpix for the WF.
\label{fig:cen_alg_47tuc_std_cat}}
\end{figure*}

Residuals from each centering algorithm's transformations are plotted in Figure \ref{fig:cen_alg_47tuc_std_cat}
as a function of $G$ magnitude.  We show
residuals along the detector $y$-axis; $x$-residuals show very similar trends. Each centering algorithm and each chip are 
specified in the respective panel. We also note the total number of stars that participated in the solution, and the standard 
error of the solution in mpix. 

As expected, the algorithms perform better with the PC than with the WF chips. This is a consequence 
of the undersampling; PC images are better sampled than WF images by a factor of $\sim 2$. The ePSF algorithm 
performs best in all four chips, however, the limits imposed on the q parameter have as a consequence the discarding of objects 
between $G = 18 -20$ mag. 
Performance of the PSFEx algorithm follows behind that of the ePSF algorithm. 
Although the standard
deviation of the PSFEx residuals appears much larger than that of the ePSF residuals 
(as seen in Fig. \ref{fig:cen_alg_47tuc_std_cat}), this is in part due to the larger magnitude range covered by the PSFEx algorithm.
For stars in common between the two, the PSFEx residuals have scatter that is from 10\% to 70\% larger, depending on the chip.
At the bright end, PSFEx effectively
outperforms the ePSF algorithm by retaining many well-measured stars, while at the faint end it outperforms the 2DG algorithm
in precision.

The TinyTim+Dolphot algorithm appears to be the most affected by undersampling. For the WF, the residuals are very scattered, and 
show a bimodal distribution, which is quite prominent at $G \le 22$. This bimodal distribution is
a result of large pixel-phase bias (see Sec. \ref{subsec:pixelphase}).

The Fourier technique, while displaying a scatter comparable to that of the 2DG algorithm, loses stars at the faint end. Also, 
chip WF2 performs more poorly than WF3 and WF4 due to the density of stars in WF2, which is closest to the cluster center. It is 
interesting that it recovers well the astrometry of bright stars almost into the saturation regime.

\begin{figure*}[h!]
\includegraphics[angle=-90,scale=0.67]{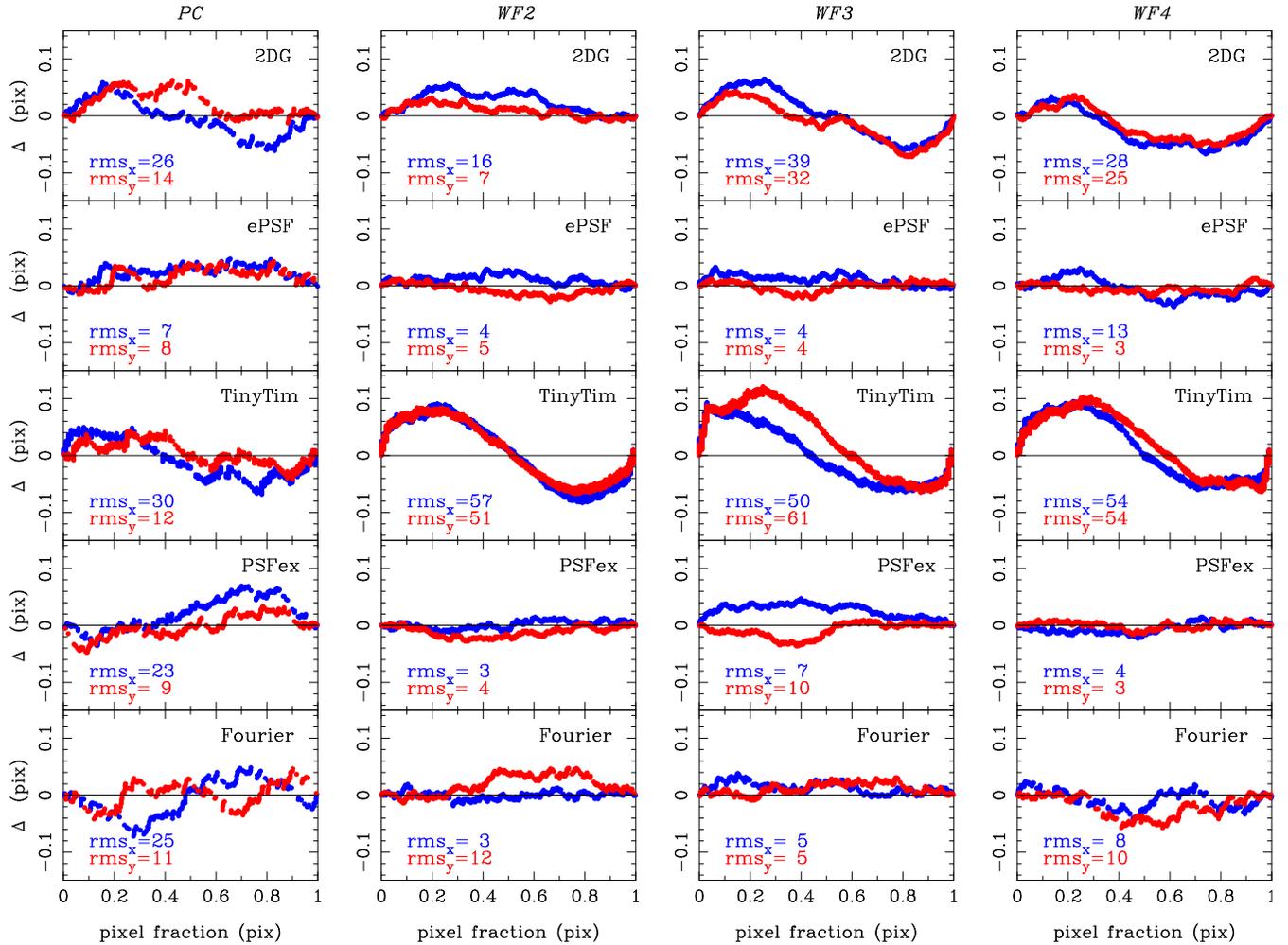}
\caption{Representation of the pixel-phase bias associated with each centering algorithm 
along the $x$ and $y$ chip axes and for each WFPC2 chip, as labeled.
Shown are differences between the measured pixel phase and that expected from a uniform
distribution, as a function of pixel phase itself (see Sec. \ref{subsec:pixelphase}).
The rms above that expected for a randomly distributed sample of the same size is also given for each
axis, in mpix.
\label{fig:pbias_alg_47tuc_std_cat}}
\end{figure*}

\subsection{Pixel-phase Bias}\label{subsec:pixelphase}
Pixel-phase bias is measured by constructing ``bias curves,'' one in each
coordinate, for a given set of image centers given in pixels.  To do so, the
stars' positions are ordered by fractional pixel phase, from 0 up to but
not including 1.  Once ordered, each star's pixel phase is compared to that
at the same rank for a set containing an equal number of stars but having a
{\it uniform} pixel-phase distribution.  The differences between measured pixel phase
and that expected for a uniform distribution, as a function of pixel phase
itself, is a representation of the pixel-phase bias curve.

For a set of centers with random (unbiased) pixel phases, such curves will be
flat, with a level of noise that depends on the number of stars in the sample.
The curves of pixel-phase biased samples will exhibit structure that is often, but not
always, roughly sinusoidal.



In Figure \ref{fig:pbias_alg_47tuc_std_cat} we plot pixel-phase bias curves
for the five centering algorithms explored here and for each chip. We use the coordinates as they are output
from the centering algorithm, i.e., with no distortion precorrection as this will alter the pixel-phases.
Only stars that participated in the solution into the standard catalog (see Sec. \ref{subsec:trans_std_catalog}) are
included in these plots.
For all algorithms the PC-curves are noisy and thus it is unclear which algorithm is best, if any.
For the WF, the TinyTim algorithm has the strongest pixel-phase bias followed by the 2DG. The ePSF, PSFEx and Fourier
algorithms have relatively flat curves. While more exploration is needed especially for the PC, we have established
that for high-precision, relative astrometry, the TinyTim and 2DG algorithms are not as desirable, while the
ePSF, PSFEx and Fourier are promising. The ePSF however, discards the bright end, while the Fourier algorithm
loses the faint end.

This, along with the relative sizes of the transformation standard errors previously discussed, lead us to adopt the
ePSF centers when maximum precision is desired, and the 2DG centers when bright-star coverage is important.

\section{Distortion Maps} \label{sec:dist_maps}
\subsection{Exploring high frequency systematics} \label{subsec:high_freq}
In Sec. \ref{sec:aims} we defined high-frequency systematics as those manifested over scales of the order or 34 pixels or 
less. The pixel-phase bias does qualify for this type of systematic. We gave a brief introduction and demonstration of this bias
in Sec. \ref{subsec:pixelphase}. A full exploration of its effect
on high-precision relative astrometry is beyond the scope of this paper and will be presented elsewhere. Importantly, its
contribution to the maps we build now will be random and at a level below that introduced by other sources such as
proper-motion errors from the standard catalogs.

In order to explore scales below 34 pixels, but larger than one pixel, we need a residual map with a high resolution, ideally of 
the order of a few pixels. In each such resolution element (or cell) we need a sufficient number of residuals to overcome the 
random errors, and furthermore require a balanced distribution of residuals over the resolution elements. In other words, if 
residuals are clumped spatially to thousands in some cells and none or a few in other cells, then the map will be 
compromised. At this resolution {\it Gaia} does not provide such a density and distribution of stars in the WFPC2 exposures.

Instead, we search for dithered WFPC2 exposure sets that are dense in stars and with small offsets between exposures. For instance,
a 47 Tuc core set (PI Gilliland, PID 8267) of 636 160-sec exposures appears to fit the bill; however its offsets are at most 
1 WF pix, meaning the stars will be clustered in certain cells of an average-residual map. Thus, to supplement this data set, 
we search for other appropriate exposures, specifically in cluster fields. 
Table \ref{tab:rel_transf_census} lists the exposure sets resulting from the search; these we use to explore high-frequency systematics. 
The aim of this exercise is to find presumed 
chip-construction-related systematics, beyond the known 34th-row issue. As such, time variations and filter differences are not 
relevant, allowing us to stack residuals from various exposure sets taken at different epochs and with different filters.

The process is as follows: detection and centering is done with the {\it HST} code using ePSFs. Pre-corrections for 34th-row 
and (nominal) optical distortion are applied. 
For each exposure set, we choose one exposure and use its set of object positions to serve as an internal
reference ``catalog.'' The set of object positions from each of the remaining exposures in this exposure group are transformed into
the reference set using third-order polynomials. Once all exposures' object lists are on the same system --- that of the reference 
exposure --- we construct an internal catalog by averaging repeated mesaures of the same object, 
with sigma-trimming to reject outliers.
Positions within a given tolerance (2 PC pix, 1 WF pix) are deemed to be of the ``same'' object.  
Each exposure's object list is then transformed into this internal, average catalog and the residuals stored and binned into cells 
based on pixel coordinates within the chip. 
Finally, for each chip, we stack residuals from all exposure sets and filters, deriving mean residual vectors on a per-cell basis,
i.e., a 2-d residual map.

Properties of the exposure sets used in our high-frequency mapping are given in Tab. \ref{tab:rel_transf_census}. 
For reference, we give the total number of residuals per set used in the map construction along with the standard deviation 
of the residual distribution for chip WF2. 
One exception is the data set of cluster NGC 6752 which was taken only with the PC; the units for this set are in PC pixels. 
The largest spatial offset within an exposure set is also listed.
For the 47 Tuc core set, which yields the most residuals, we use only every 10th residual available, so as not to allow
this set of exposures to completely dominate the map.

\begin{table*}
\caption{List of exposure sets used for the high-resolution map }
\begin{tabular}{rlrrrrrrrr}
\hline
\multicolumn{1}{c}{Filter} & \multicolumn{1}{c}{Target} & \multicolumn{1}{c}{N$_{resid}$} & \multicolumn{1}{c}{$\sigma_{x}$}  & 
\multicolumn{1}{c}{$\sigma_{y}$} & \multicolumn{1}{c}{N$_{exp}$}
& \multicolumn{1}{c}{Offset} & \multicolumn{1}{c}{Exp. time} & \multicolumn{1}{c}{Epoch} &
\multicolumn{1}{c}{PID} \\
 & & & \multicolumn{2}{c}{(WF mpix)} & & \multicolumn{1}{c}{(WF pix)} & (sec) &  & \\
\hline
555 & 47 Tuc core$^1$ & 170020 & 19 & 18 & 636 & 1 & 160         & 1999.5 & 8267 \\
555 & ngc 6341 & 334582 & 46 & 48 & 128 & 176 & 10,100,400       & 2008.1 & 11077 \\
555 & ngc 6441 & 154052 & 31 & 35 & 36 & 4 & 160                 & 2007.3 & 10474 \\
555 & {\it ngc 6752$^2$} & 309365 & {\it 27} & {\it 30} & 162 & {\it 563} & 2,26,80      & 1994.6 & 5318 \\
\hline
606 & ngc 6397        & 68625 & 20 & 18 & 121 & 7  & 500,600     & 2005.2 & 10424   \\
606 & ngc 6656 group1 & 47494 & 14 & 14 & 24 & 16 & 260          & 1999.1 & 7615 \\
606 & ngc 6656 group2 & 57619 & 15 & 16 & 24 & 16 & 260          & 1999.1 & 7615 \\
606 & ngc 6656 group3 & 59351 & 17 & 18 & 24 & 16 & 260          & 1999.1 & 7615 \\
606 & ngc 7078        & 38422 & 42 & 44 & 33 &  9 & 300,500      & 2001.4 & 9244  \\
\hline
814 & 47 Tuc core$^1$ & 195072   & 19 & 16 & 653 & 1 & 160       & 1999.5 & 8267 \\
814 & ngc 5139    & 72232    & 28 & 27 & 21 & 12 & 80            & 2008.1 & 11030 \\
814 & ngc 6121    & 37048    & 34 & 39 & 166 & 23 & 600,700      & 1995.2 & 5461 \\
814 & ngc 6205    & 88870    & 20 & 18 & 30 & 26 & 23,140        & 1999.9 & 8278 \\
814 & ngc 6397    & 88922    & 21 & 19 & 126 & 24 & 600          & 2005.2 & 10424 \\
814 & ngc 6656 group1    & 162917    & 15 & 15 & 64 & 7 & 260    & 1999.1 & 7615 \\
814 & ngc 6656 group2    & 182913    & 20 & 20 & 64 & 8 & 260    & 1999.1 & 7615 \\
814 & ngc 6656 group3    & 180089    & 19 & 18 & 64 & 6 & 260    & 1999.1 & 7615 \\
814 & {\it ngc 6752$^2$}    & 281132  & {\it 20} & {\it 20} & 146 & {\it 562} &  50,160  & 1994.6 & 5318 \\
\hline
\multicolumn{10}{l}{1 - used only every tenth residual} \\
\multicolumn{10}{l}{2 - only PC exposures; units are PC pixels} \\
\end{tabular}
\label{tab:rel_transf_census}
\end{table*}

\begin{figure*}[!h]
\includegraphics[angle=-90,scale=0.8]{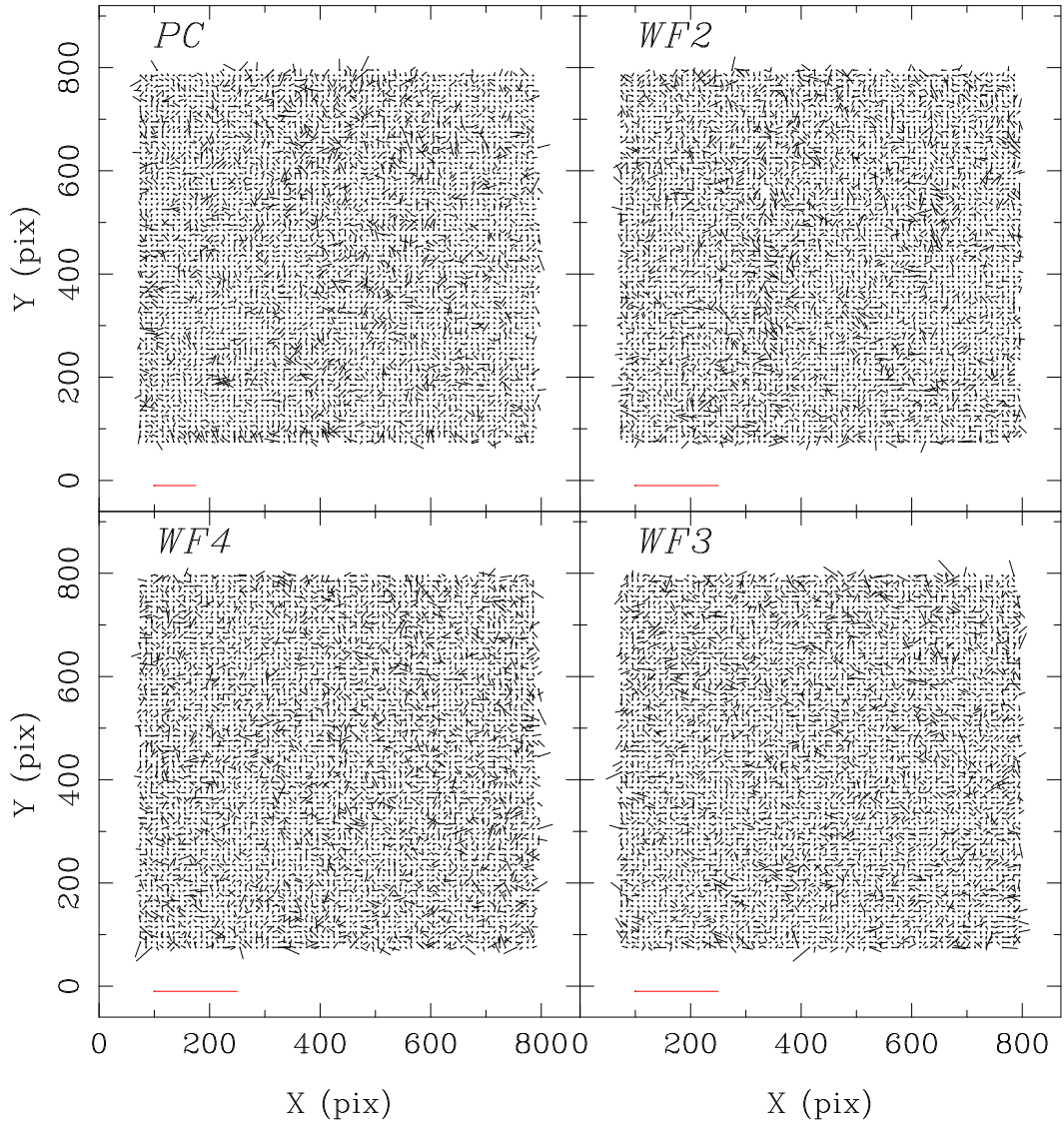}
\caption{Map of residuals with 10-pixel resolution element. Any resolution element that had an average residual
  $|r| > 20$ mpix for the PC and $|r| > 10$ mpix for the WF was excluded from the map, thus eliminating some outlier resolution 
elements. At the edges, resolution elements with large residuals still can be found. However, no regular, distinct pattern can be 
discerned from this map. The horizontal line in the bottom left of each panel represents a vector of 50 mpix size.
\label{fig:map_10_7}}
\end{figure*}

A map is built by specifying the cell size (i.e., resolution), a smoothing radius, and a maximum allowed residual amplitude, so
as to eliminate outliers. Within each cell an average residual vector is calculated (we opt to use a sigma-trimmed mean) along
with its statistical uncertainty, i.e.,
the standard deviation divided by the square root of the number of residuals in the cell. 
We discard residuals exceeding 200 mpix (in both PC and WF) to avoid undue influence by possible outliers. 
Two sets of maps are constructed, one with a cell dimension of 10 pixels and smoothing radius of 7 pixels, and a second set
using a cell size of 6 pixels and smoothing radius 4 pixels.
The 10-pixel resolution map has of the order of 560 residuals per resolution element. The formal error of the computed mean
residuals is on average 1.5 mpix for the PC and 1.3 to 1.6 mpix for the WF chips.
The 6-pixel resolution map has of the order of 190 residuals per resolution element.
The formal error of the computed mean residuals is on average 2.7 mpix for the PC and 2.3 to 2.6 mpix for the WF chips.
Figure \ref{fig:map_10_7} plots the mean residual vectors of the 10-pixel resolution map. Any resolution element that had an 
average residual
$|r| > 20$ mpix for the PC and $|r| > 10$ mpix for the WF was excluded from the plot, as these were determined to be outlier values.
Beyond the 34-row problem, there is no indication of any additional systematic pattern related to the chip manufacturing process,
such as was found for the WF3 and ACS cameras \citep{koz16}. 
Inspection of the 6-pixel resolution map, although noisier than the 10-pixel map, also reveals no discernible systematic pattern.

\subsection{Exploring low frequency systematics} \label{subsec:low_freq}
\subsubsection{Map construction} \label{subsubsec:build}
Systematics with a spatial extent larger than 34 pixels would most likely be due to uncorrected optical distortion.
Therefore, separate maps are constructed for each filter. Residual maps based on each of the three, external
standard catalogs listed in Tab. \ref{tab:std_cat} will be examined.
The $\omega$ Cen-center catalog has overlap only with F555W exposures, while the 47 Tuc off-center and {\it Gaia} EDR3 catalogs have 
overlap with exposures in all three filters. We explore residuals generated from both 2DG and ePSF positions.

Pixel coordinates are pre-corrected for the 34th row and nominal cubic distortion. For each observation and chip, external catalog
coordinates are transformed into chip coordinates and a cubic solution is performed between these and the pixel coordinates of the 
WFPC2 exposure. Solutions into {\it Gaia} EDR3 are retained only when all three of the following conditions are met: 
the standard error of the solution in both coordinates is less than 0.7 pix for the PC and 1.5 pix in the WF, 
the scale term in both coordinates deviates by less than $10\%$ from its nominal value, 
and the number of reference stars in the solution is larger than 20 for the PC and larger than 30 for the WF chips.
Solutions that do not meet these requirements are discarded from the analysis. A similar scheme is adopted for 
solutions into the other two standard catalogs, always verifying the quality of the solution.

Finally, residuals are stacked and averaged by resolution element, as before, producing a residual map. 
For {\it Gaia} EDR3 residuals, we also force each resolution element of 
the map to not have more than 4 residuals of the same {\it Gaia} star. This minimizes the contribution of any poorly-measured {\it 
Gaia} stars to the average for any given resolution element. We also restrict residuals to be only from {\it Gaia} stars with $G 
= 11 - 20$, and discard outlier residuals, i.e. with
absolute values larger than 2 pix for the PC and 1 pix for the WF.

For the low-frequency maps a 50-pix resolution element is used and a smoothing radius of 50 pix. 
The standard error of the average is calculated for each resolution element,
as the standard deviation of the residuals divided by the square root of the number in the cell. 
Table \ref{tab:error_res_elem} lists the average properties of the derived maps for each catalog, filter, and centering 
algorithm. 
The median standard error, in mpix, per resolution element is also listed, as is
the number of residuals per resolution element. 
For the standard error, we give
only the x-coordinate values, with y-coordinate results having similar values. 
Also listed is the total number of residuals that participated in
the construction of each map. The standard error per resolution element is the primary indicator of map quality.
The {\it Gaia } EDR3 maps supersede our initial {\it Gaia } DR2-based maps. The DR2 maps had median errors that are a factor 
of 1.4 to 1.5 larger than those of the EDR3 maps, due to the DR2 catalog's less precise
proper motions as well as it having a smaller number of stars available for the solutions.

Inspecting the {\it Gaia } EDR3 map properties in Tab. \ref{tab:error_res_elem}, one sees that maps using the 2DG
centering algorithm have better standard errors than those using the ePSF algorithm; this is simply due to the
substantially larger number of residuals that participated in the construction of the 2DG maps, consistently across
all three filters. It is also an indication that the dominant source of error in the map is due to proper-motion errors in {\it 
Gaia } EDR3, i.e., differences in the
errors associated with the centering algorithms are not discernible in this circumstance.

This is not the case for the special catalogs; ePSF-based maps are slightly better than 2DG ones, while the
numbers of residuals in the two algorithms' maps are comparable. Thus, the precision of the proper motions
of these catalogs allows for the centering precision's contribution to the total error budget to become apparent.
This is particularly obvious for the 47 Tuc catalog maps, while being somewhat more ambiguous for the $\omega$ Cen catalog.

Based on the standard error estimates in Tab. \ref{tab:error_res_elem}, the overall best maps would seem to be
those using {\it Gaia } EDR3 with 2DG centering, although the pairing of the other two standard catalogs with ePSF centering 
follow close behind. 
In fact, the latter combination appears to be the better choice for the PC in all filters. 
With this in mind, we will pursue to construct weighted averages of the maps derived from {\it Gaia } EDR3 using 2DG centering 
along with maps derived using the two special standard catalogs and ePSF centering.

\begin{table}
\caption{Median error (in mpix) per resolution element in the x-coordinate for the 50-pix resolution map }
\begin{tabular}{rrrr|rrr}
\hline
 & \multicolumn{3}{c}{2DG} & \multicolumn{3}{c}{ePSF} \\
& \multicolumn{1}{c}{error} & \multicolumn{1}{c}{N/cell} & \multicolumn{1}{c}{N$_{all}$} & \multicolumn{1}{c}{ error} & 
\multicolumn{1}{c}{N/cell} & \multicolumn{1}{c}{N$_{all}$} \\
\hline
\multicolumn{7}{c}{{\it Gaia} EDR3} \\
\hline
\hline
\multicolumn{7}{l}{\bf \it F555W} \\
PC &  9.6 & 2169 & 143120  &  11.2 & 1730 &      113367 \\
WF2 & 2.0 & 6416 & 441392  &  2.4 & 3851 &       271620 \\
WF3 & 2.0 & 6169 & 416067  &  2.5 & 3631 &       254464 \\
WF4 & 1.9 & 6534 & 440822  &  2.2 & 4030 &       279304 \\
\multicolumn{7}{l}{\bf \it F606W} \\
PC & 9.5 & 961 & 63347 & 15.0 & 529 &        54188 \\
WF2 & 1.7 & 4976 & 323628 & 3.1 & 1563 &    290753 \\
WF3 & 1.7 & 4474 & 323285 & 3.2 & 1610 &    290193 \\
WF4 & 1.6 & 5058 & 328403 & 2.9 & 1725 &    295793 \\
\multicolumn{7}{l}{\bf \it F814W} \\
PC & 10.2 & 1771 & 116730 & 12.1 & 1405 &      93192 \\
WF2 & 1.8 & 6156 & 405651 & 2.3 & 3470 &     345662 \\
WF3 & 1.9 & 5991 & 394326 & 2.4 & 3696 &     340264 \\
WF4 & 1.8 & 6281 & 414623 & 2.3 & 3895 &     354668 \\
\hline
\multicolumn{7}{c}{47 Tuc - off center} \\
\hline
\hline
\multicolumn{7}{l}{\bf \it F555W} \\
PC & 6.2 & 115 & 9611 & 5.4 & 108 &    9197 \\
WF2 & 2.9 & 480 & 34386 & 1.8 & 474 &   33119 \\
WF3 & 4.2 & 265 & 18463 & 3.2 & 235 &   17518 \\
WF4 & 4.3 & 261 & 18520 & 3.1 & 262 &   18851 \\
\multicolumn{7}{l}{\bf \it F606W} \\
PC & 2.6 & 325 & 22743 & 2.0 & 305 &    21090 \\
WF2 & 1.7 & 1310 & 86418 & 0.9 & 1195 &   80089 \\
WF3 & 1.6 & 1280 & 84536 & 1.0 & 1199 &   80000 \\
WF4 & 1.8 & 1249 & 83027 & 0.9 & 1174 &   78510 \\
\multicolumn{7}{l}{\bf \it F814W} \\
PC & 3.8 & 129 & 9038 & 3.5 & 125 &    8844 \\
WF2 & 2.2 & 682 & 46208 & 1.5 & 646 &   44053 \\
WF3 & 2.5 & 558 & 37329 & 1.9 & 527 &   36074 \\
WF4 & 2.8 & 456 & 31489 & 1.9 & 444 &   30771 \\
\hline
\multicolumn{7}{c}{$\omega$ Cen - center} \\
\hline
\hline
\multicolumn{7}{l}{\bf \it F555W} \\
PC & 6.2 & 269 & 17957 & 4.0 & 266 &    17761 \\
WF2 & 5.1 & 363 & 26334 & 2.8 & 453 &   32766 \\
WF3 & 4.8 & 473 & 30961 & 2.9 & 555 &   37262 \\
WF4 & 4.1 & 588 & 39840 & 2.3 & 726 &   49725 \\
\hline
\end{tabular}
\label{tab:error_res_elem}
\end{table}


\subsubsection{Centering-Algorithm Field Effects} \label{subsubsec:cen_eff}
Before spatial maps based on different centering algorithms can be combined, it must be demonstrated that differences between 
the two types of centers themselves do not introduce a spatial variation.
To check for this, we select roughly 10\% of all WFPC2 exposures in the three filters and, for each, transform 2DG positions of
anonymous stars into ePSF positions of the same stars using general cubic terms, similar to Eq.~1 but using raw positions.
Choosing the best such solutions -- those with at least 200 stars in common and standard errors less than 80 mpix --
residuals are accumulated per chip and filter.
From these we construct low-frequency residual maps as before, but in this case the maps 
represent any artificial field dependence due to the differences in centering method.

These 2DG-into-ePSF maps do show significant structure, in all chips and filters.
A measure of the significance of a map is its mean $\chi^2$, \

\setcounter{equation}{1}\label{chisq}
\begin{equation}
<\chi^2> = \sum (r/\sigma)^2 / N 
\end{equation}
where $r$ is the signal in a map element, $\sigma$ its uncertainty, and $N$ is the total number of elements.

The 2DG-into-ePSF maps have $<\chi^2>$ values of $\sim3$ for the PC, and range from 4 to 14 for the WF chips, 
the largest values being found for filter F606W.
Thus, 2DG-based maps {\it cannot} be combined with ePSF-based maps, unless the former are somehow first put onto the ``system'' of the latter.
Fortunately, the just-constructed 2DG-into-ePSF maps can be used to accomplish exactly this; a simple subtraction of the 
appropriate 2DG-into-ePSF map from any 2DG-based map will adjust it to the ePSF system of centering.
Due to the large number of residuals used to build them, subtracting the 2DG-into-ePSF map
adds negligibly to the target-map uncertainties.

From this point on, all 2DG-based maps are adjusted to the ePSF-centering system upon their construction.
To emphasize their modified nature, we will refer to such maps (and cubic terms) as 2DG$^*$.

In addition to the high-order field effects induced by the different centering schemes, we point out that the cubic terms are also affected.
As such, we calculate the mean cubic terms in the 2DG-into-ePSF solutions and use these as adjustments to any standard-catalog derived
2DG-based cubic terms. 
The resulting 2DG$^*$ cubic terms are then applicable to ePSF-determined positions.

We have chosen to work in the system of ePSF centering because it is, without doubt,
the more common standard of the two 
for astrometric processing of WFPC2 images.

\subsubsection{Remaining Map Differences} \label{subsubsec:map_diff}
Differences between maps derived from standard catalogs are examined to determine if they are significant, with respect to the
uncertainties of the standard catalogs and the centering precision of the WFPC2 images.
For each resolution element we calculate a
difference divided by its estimated formal error. If the difference in maps is merely random, the
normalized differences should distribute around zero with a standard deviation of 1.
We determine the standard deviation of the normalized differences $\sigma_s$ 
using probability plots \citep{ham78}
with trimming of the extreme $10\%$ of both wings to alleviate the effect of possible outliers.\footnote{\label{hamaker}The probability-plot method is a robust estimator of mean and standard deviation of a
supposed Gaussian distribution of values. Comparing the value-ordered spacing of the central portion of the set of values
with that expected for a similarly sized Gaussian sample renders it less susceptible to spurious outlier points in the wings.}
To demonstrate, we apply the procedure to maps based on
the same standard catalog, centering algorithm, and filter and consider differences
between chip WF2 and the other three chips. 
The calculated map-difference dispersions have values around 2.0.
The smoothing used during the construction of these low-frequency maps effectively decreases the
number of independent resolution elements by a factor of $\pi$, from $15\times15$ to $\sim 72$.  
Thus, the uncertainty in a dispersion measure of 2.0 will be $\sim 0.2$,
indicating 
that indeed the map of WF2 is significantly different from the maps of the other three chips.

\begin{figure}
\includegraphics[angle=0,scale=0.47]{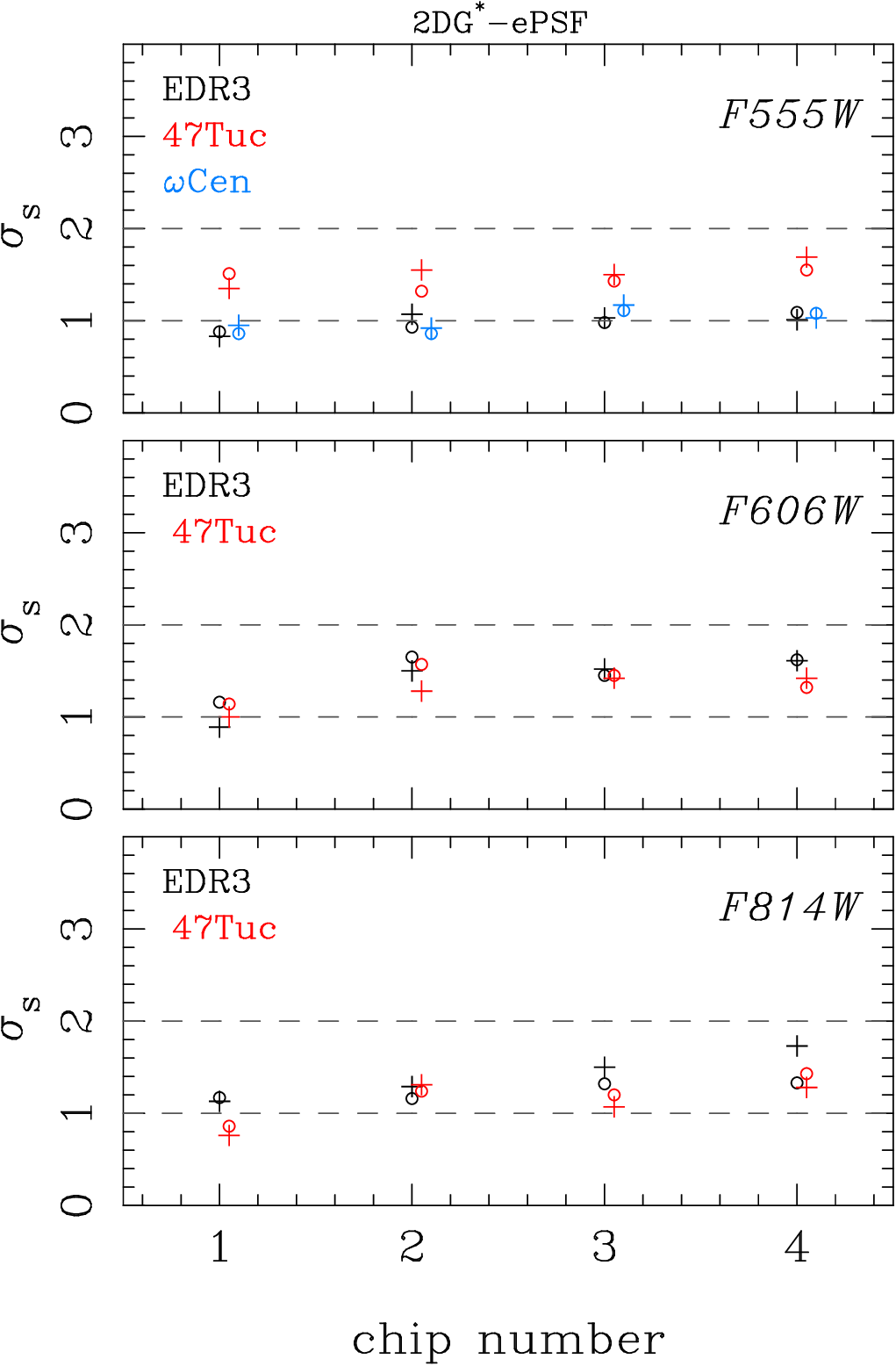}
\caption{Dispersions of the error-normalized differences comparing maps derived from standard catalogs, based on 2DG$^*$ and ePSF centers. 
  Color-coding, as labeled, indicates the standard catalog used in constructing the maps being compared. 
  Each panel presents results for a different filter. 
  The dispersion statistic for differences between x-coordinate maps are plotted with
  circles, results for y-coordinate maps are represented with plus signs.
\label{fig:mapdiff_calg}}
\end{figure}
\begin{figure}
\includegraphics[angle=0,scale=0.40]{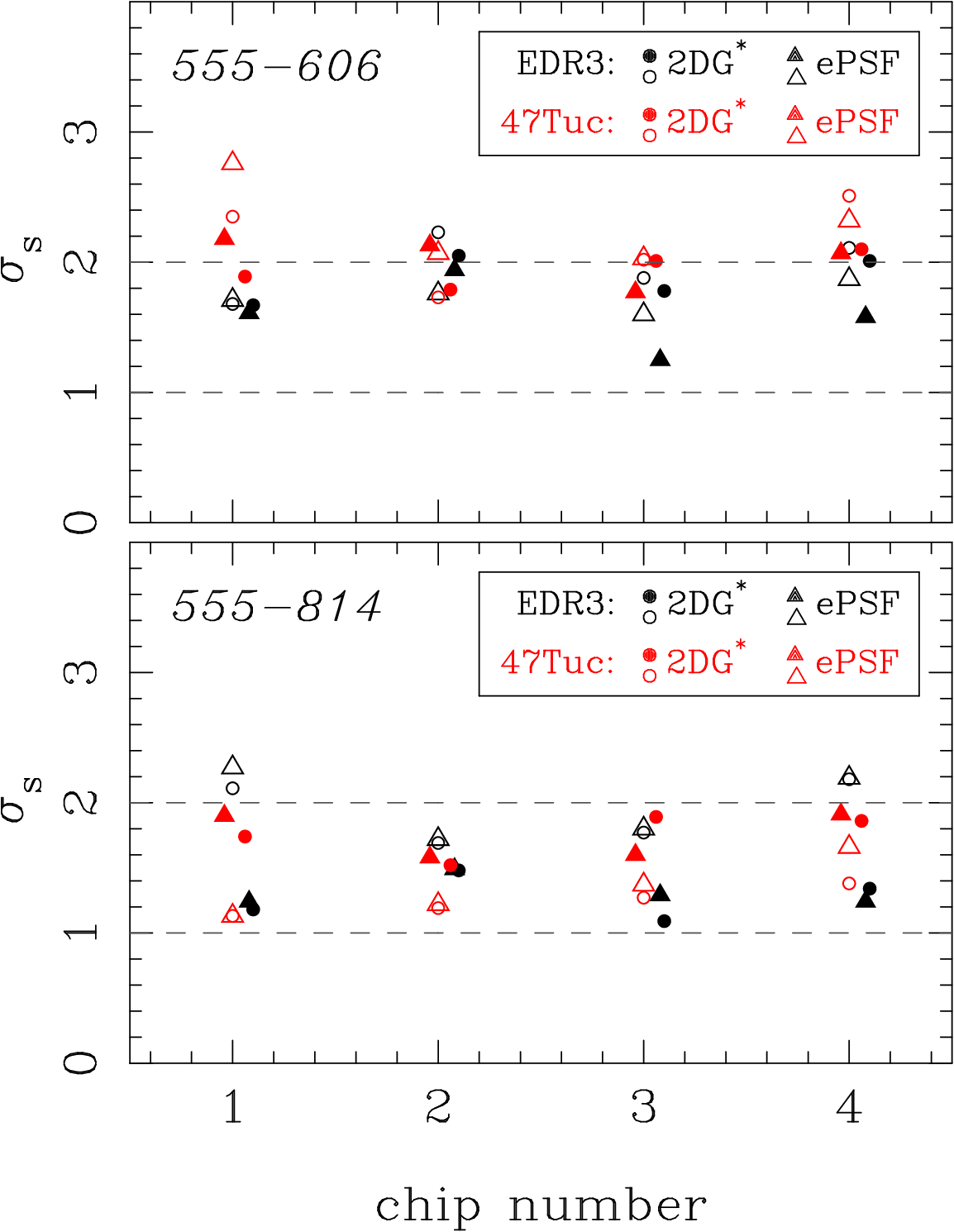}
\caption{Dispersions of the error-normalized differences comparing maps derived in different filters.
  The top panel shows
  statistics for map differences between F555W and F606W; the bottom panel compares F555W and F814W maps.
  Comparisons of EDR3-based maps are shown with black symbols, while the 47Tuc-based map differences are shown in red.
  Differences for 2DG$^*$-algorithm maps are represented with circles, while the ePSF maps are represented with triangles. Filled
  symbols depict x-coordinate map results, open symbols depict y-coordinate results.
\label{fig:mapdiff_filter}}
\end{figure}

Figure \ref{fig:mapdiff_calg} shows this statistic ($\sigma_s$) for differences between maps
based on the 2DG$^*$ and the ePSF algorithms, i.e., after having adjusted the 2DG maps as described in Sec.~7.2.2.  
For filters F555W and F814W the dispersion of the normalized differences $\sigma_s$ has values close to $1$, indicating that differences in 
centering algorithms are not discernable in these maps. 
For filter F606W, $\sigma_s$ has values closer to 1.5 for the wide-field chips, hinting at some remaining difference between 
centering algorithms.  
This is possibly related to the use of the F555W effective-psf library for the ePSF centering of the F606W images, 
the best available option at this time. 
Our overall conclusion is that centering algorithm differences are not appreciably affecting the distortion maps, 
provided the 2DG-center maps are properly adjusted.
This will allow us to average maps based on the two different centering schemes.

In Figure \ref{fig:mapdiff_filter} we show $\sigma_s$ for map differences between filters. 
Values of $\sigma_s \sim  2$ are obtained for differences between F555W and F606W indicating that the
distortion maps are different for these filters. The difference between F555W and F814W is less striking with
$\sigma_s \sim 1.5 $. 
In general, filter differences do appear significant and thus we will provide separate maps for each filter.

\section{Adopted Distortion Model} \label{sec:dist_model}
Corrections to the nominal distortion model that we present consist of a) third-order terms based on EDR3 solutions of 2DG$^*$ positions,
i.e., the terms are adjusted to be applicable to ePSF-determined positions.
and b) higher than third order residual maps based on a combination of EDR3, 47 Tuc and $\omega$ Cen standard catalogs.
The coefficients of the third order terms are given at two average epochs: an early epoch ($< 2005.0$) and a late epoch ($\ge 2005.0$).
EDR3 solutions by far outnumber solutions from the other standard catalogs (see Tab. \ref{tab:std_cat}).
Therefore we determine each coefficient
by averaging solely the EDR3 solutions for each chip, filter and epoch group.
In Table \ref{tab:third_order_terms} we list these coefficients together with the average epoch.
When calculating the correction for a given observation we choose to use a linear
interpolation/extrapolation between these two mean epochs to determine the applicable cubic coefficients.
Alternatively, the user can decide in which epoch group the observation is, and directly use the coefficients listed
in Tab. \ref{tab:third_order_terms}. Variations with time are mild, nevertheless
significant in some cases (see Tab. \ref{tab:third_order_terms}).

\begin{table*}
\caption{Average cubic terms of the EDR3 2DG$^*$ solutions, (on the ePSF system). Units are $10^{-11}$ pix/pix$^3$}
\begin{tabular}{lrrrr|rrrr}
\hline
\multicolumn{1}{c}{$<epoch>$} 
 & \multicolumn{1}{c}{XXX} & \multicolumn{1}{c}{XXY} & \multicolumn{1}{c}{XYY} & \multicolumn{1}{c}{YYY}
 & \multicolumn{1}{c}{YYY} & \multicolumn{1}{c}{YYX} & \multicolumn{1}{c}{YXX} & \multicolumn{1}{c}{XXX} \\
 \hline
 \multicolumn{5}{c}{\it F555W --- X-sol.} & 
 \multicolumn{4}{c}{\it F555W --- Y-sol.} \\
  {\bf PC } \\
1997.76 & $-039(028)$  & $ 078(027)$  & $-158(025)$  & $-019(029)$  & $-009(027)$  & $-050(027)$  & $-136(026)$  & $ 008(024)$  \\
2008.05 & $-135(025)$  & $ 136(028)$  & $-096(023)$  & $-098(026)$  & $-182(022)$  & $ 070(024)$  & $ 103(025)$  & $ 176(023)$  \\
  {\bf WF2} \\
1998.15 & $ 013(006)$  & $-049(005)$  & $-013(005)$  & $-043(005)$  & $-011(006)$  & $ 010(005)$  & $-038(006)$  & $-011(006)$  \\
2007.70 & $-006(006)$  & $-034(006)$  & $-014(006)$  & $-069(007)$  & $-002(007)$  & $ 012(007)$  & $-037(007)$  & $ 008(007)$  \\
  {\bf WF3} \\
1998.13 & $-001(006)$  & $ 019(005)$  & $ 019(006)$  & $-031(006)$  & $-005(006)$  & $ 028(006)$  & $ 007(006)$  & $-004(006)$  \\
2007.66 & $ 018(006)$  & $ 005(006)$  & $ 010(005)$  & $-029(005)$  & $-011(006)$  & $-014(005)$  & $ 041(006)$  & $ 002(007)$  \\
  {\bf WF4} \\
1998.15 & $ 005(006)$  & $-015(005)$  & $-022(005)$  & $-039(005)$  & $ 026(006)$  & $-051(006)$  & $-007(005)$  & $ 005(006)$  \\
2007.71 & $-013(006)$  & $-009(006)$  & $-018(006)$  & $-074(006)$  & $-025(007)$  & $-025(006)$  & $ 003(006)$  & $ 036(007)$  \\
 \hline
 \multicolumn{5}{c}{\it F606W --- X-sol.} & 
 \multicolumn{4}{c}{\it F606W --- Y-sol.} \\
  {\bf PC } \\
2000.88 & $-079(027)$  & $-021(027)$  & $-064(028)$  & $ 158(029)$  & $-097(028)$  & $-168(034)$  & $ 081(028)$  & $ 102(026)$  \\
2007.08 & $-317(055)$  & $-196(047)$  & $ 026(044)$  & $-052(038)$  & $-076(039)$  & $-005(032)$  & $ 066(030)$  & $-061(038)$  \\
  {\bf WF2} \\
2000.97 & $ 047(005)$  & $-006(005)$  & $-015(005)$  & $-006(005)$  & $-017(006)$  & $ 021(005)$  & $ 009(005)$  & $-008(005)$  \\
2007.77 & $-007(010)$  & $-009(009)$  & $-045(009)$  & $ 007(008)$  & $-056(011)$  & $ 021(009)$  & $-038(011)$  & $-002(009)$  \\
  {\bf WF3} \\
2000.96 & $ 005(005)$  & $ 017(005)$  & $ 019(005)$  & $-028(005)$  & $ 040(006)$  & $ 003(005)$  & $-036(006)$  & $ 037(005)$  \\
2007.64 & $-021(008)$  & $ 066(008)$  & $ 013(008)$  & $-056(008)$  & $ 042(009)$  & $-008(010)$  & $-042(008)$  & $ 017(008)$  \\
  {\bf WF4} \\
2000.95 & $ 013(005)$  & $-030(005)$  & $-013(005)$  & $-032(004)$  & $ 024(005)$  & $-069(005)$  & $-006(006)$  & $-012(006)$  \\
2007.65 & $ 026(009)$  & $-064(008)$  & $-024(007)$  & $-017(008)$  & $ 016(008)$  & $-055(007)$  & $-006(010)$  & $-021(009)$  \\
 \hline
 \multicolumn{5}{c}{\it F814W --- X-sol.} & 
 \multicolumn{4}{c}{\it F814W --- Y-sol.} \\
  {\bf PC } \\
1998.39 & $ 002(021)$  & $ 014(023)$  & $-116(022)$  & $ 102(024)$  & $-025(024)$  & $-094(023)$  & $-016(022)$  & $ 029(020)$  \\
2007.12 & $-112(033)$  & $-100(030)$  & $-097(029)$  & $-038(026)$  & $ 022(026)$  & $ 004(026)$  & $ 125(027)$  & $ 024(026)$  \\
  {\bf WF2} \\
1998.69 & $ 050(004)$  & $-077(004)$  & $ 038(005)$  & $-014(004)$  & $ 006(005)$  & $-022(005)$  & $ 070(005)$  & $-024(005)$  \\
2007.16 & $ 028(006)$  & $-066(006)$  & $ 023(005)$  & $-021(007)$  & $-007(006)$  & $-003(005)$  & $ 042(007)$  & $ 001(007)$  \\
  {\bf WF3} \\
1998.63 & $ 033(005)$  & $ 008(004)$  & $ 045(005)$  & $-025(005)$  & $ 016(005)$  & $ 002(004)$  & $ 053(005)$  & $ 015(005)$  \\
2007.11 & $ 021(006)$  & $ 022(006)$  & $ 057(005)$  & $-035(006)$  & $ 045(007)$  & $-027(005)$  & $ 021(006)$  & $ 001(007)$  \\
  {\bf WF4} \\
1998.62 & $ 003(004)$  & $-063(004)$  & $ 063(004)$  & $-020(004)$  & $ 046(005)$  & $-102(004)$  & $ 075(004)$  & $-007(005)$  \\
2007.12 & $ 034(006)$  & $-066(005)$  & $ 060(006)$  & $-014(005)$  & $ 038(007)$  & $-085(006)$  & $ 130(007)$  & $-007(006)$  \\

\hline
\end{tabular}
\label{tab:third_order_terms}
\end{table*}

Higher than third-order distortion residual maps are built at a resolution of 50 pixel and at single epoch.
We have explored time variations in the distortion maps. Unfortunately, for the late epoch ($> 2005$) there are few
WFPC2 observations compared to earlier data, and thus late-epoch maps are rather noisy due to the scarcity of residuals.
To this end, we could not convincingly find a time dependence of these maps.

\begin{figure*}
\includegraphics[angle=-90,scale=0.67]{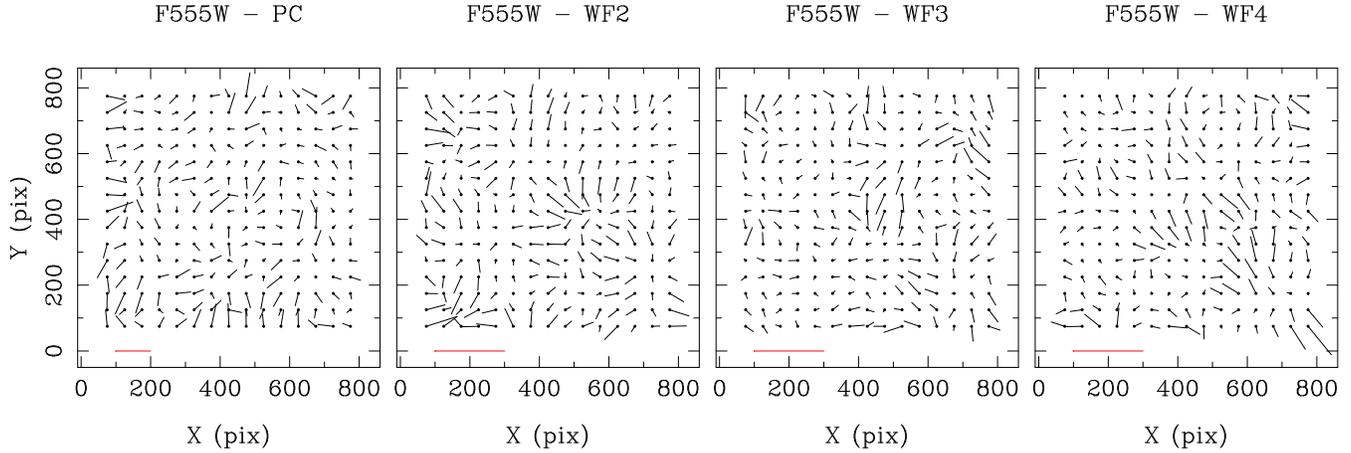}
\caption{Distortion-correction maps, post third-order correction.
  All maps have a resolution of 50 pixels, and are constructed from a weighted
  average of EDR3, 47 Tuc and $\omega$ Cen standard catalogs (see text).
  Here we show maps for filter F555W. The origin of each vector
  is marked with a black dot. The horizontal line at the bottom left
  of each plot represents a vector of 50 mpix size.
\label{fig:f555w_map}}
\end{figure*}

\begin{figure*}
\includegraphics[angle=-90,scale=0.67]{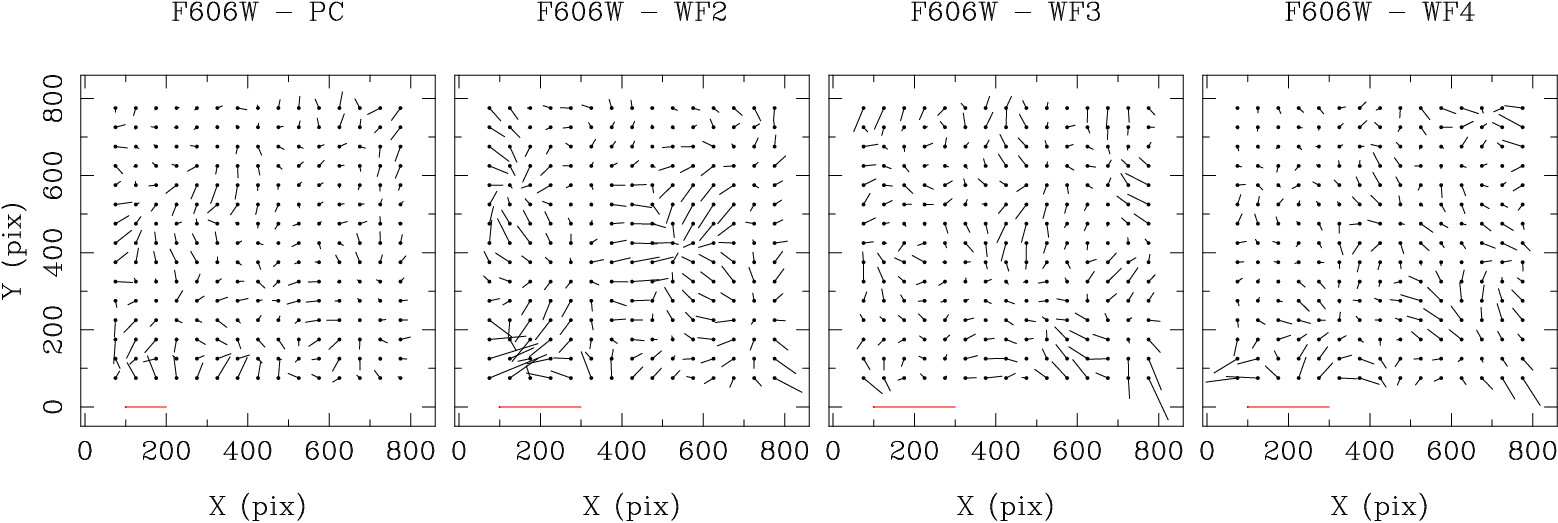}
\caption{Same as in Fig. \ref{fig:f555w_map} only for filter F606W.
  Again, the horizontal line at the bottom left
  of each plot represents a vector of 50 mpix size.
\label{fig:f606w_map}}
\end{figure*}

\begin{figure*}
\includegraphics[angle=-90,scale=0.67]{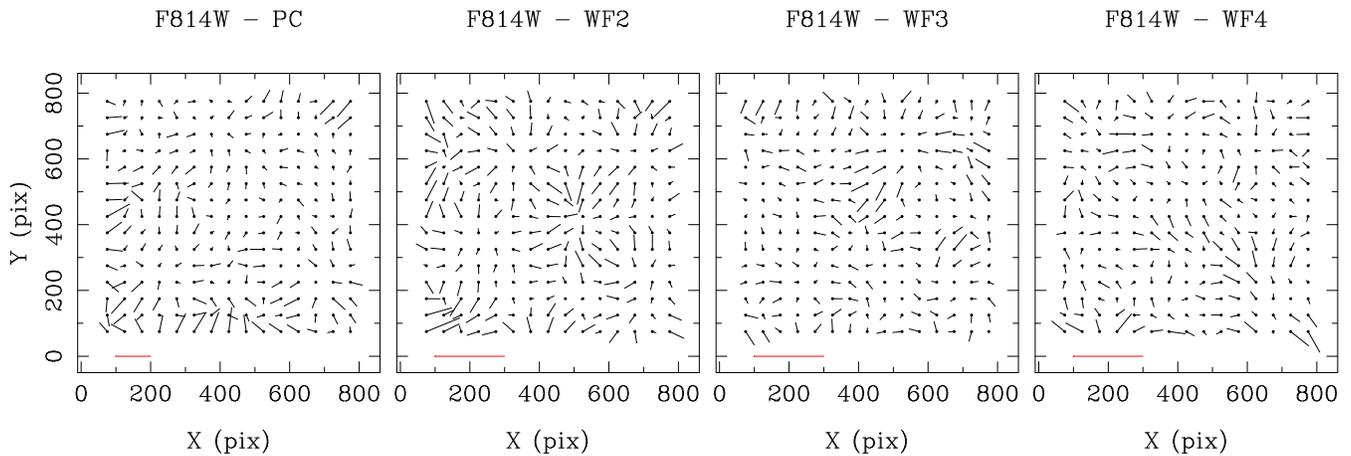}
\caption{Same as in Fig. \ref{fig:f555w_map} only for filter F814W.
  The horizontal line at the bottom left
  of each plot represents a vector of 50 mpix size.
\label{fig:f814w_map}}
\end{figure*}

For the final maps we combine residuals from the following solutions: EDR3 with 2DG$^*$ positions,
47 Tuc with ePSF positions and $\omega$ Cen with ePSF positions. According to Tab.  \ref{tab:error_res_elem}, these
solutions have the best meadian error per resolution element. We adopt a weighted mean,
where the weight is determined from the estimated uncertainty of each resolution element 
which in turn was based on the scatter and number of residuals per element.
Maps are constructed for each chip and filter;
these are presented in Figures \ref{fig:f555w_map}, \ref{fig:f606w_map} and 
\ref{fig:f814w_map}. Higher-order systematics are readily apparent in the maps.  
The $<\chi^2> $ (see eq. 2) for these maps ranges between 7 and 70.



The corrections derived here represent small post-corrections to the nominal third-order distortion determined by AK03. 
For instance, the nominal AK03 cubic-distortion corrections exhibit an amplitude
of up to 4 pixels from chip center to edge (both PC and WF). 
Inspection of the newly determined (post-nominal correction) cubic terms in Tab. 
\ref{tab:third_order_terms}, shows
that the largest terms are $\sim 300\times10^{-11}$  pix/pix$^3$ for the PC, and $\sim 100\times10^{-11}$  pix/pix$^3$ for the WF. 
These corrections are on the order of 0.1 pixels, chip center to edge.

To better compare the relative size of these corrections, we generate 10,000 randomly located points over the chip
and calculate at each location the
size of the distortion as given by AK03, by our third-order correction, and by higher than third order corrections.
Epoch is also randomly assigned for the purpose of calculating our third-order corrections.
We calculate the rms values, $\sqrt{(dx^2+dy^2)}$, where $dx$ and $dy$ are the correction values in each axis.
We tabulate these rms values for each chip and filter in Table \ref{tab:rms_corr}.
Our third-order correction is at a level of a few percent that from AK03 for the PC, and around 1\% for the WF. 
The higher-order corrections are between $\sim 18\%$ and 90\% the size of our cubic-term correction.

\begin{table}
  \caption{RMS values of various distortion corrections, and their ratios.
    Correction types are: $\delta1$ for AK03, $\delta2$ for our third-order corrections, 
    and $\delta3$ for the higher-order distortion maps.}
\begin{tabular}{rrrrrrr}
  \hline
  \multicolumn{1}{c}{Filter} & 
  \multicolumn{1}{c}{$\delta1$} & 
  \multicolumn{1}{c}{$\delta2$} & 
  \multicolumn{1}{c}{$\delta3$} &
  \multicolumn{1}{c}{$\delta2$/$\delta1$} &  
  \multicolumn{1}{c}{$\delta3$/$\delta1$} &  
  \multicolumn{1}{c}{$\delta3$/$\delta2$} \\
  \multicolumn{1}{c}{:chip} & 
  \multicolumn{1}{c}{(mpix)} & 
  \multicolumn{1}{c}{(mpix)} &
  \multicolumn{1}{c}{(mpix)} &
  \multicolumn{1}{c}{($\%$)} & 
  \multicolumn{1}{c}{($\%$)} & 
  \multicolumn{1}{c}{($\%$)} \\
  \hline
  555:1 & 1580.2 & 46.0 & 14.1 & 2.9 & 0.9 & 30.6 \\
  606:1 & 1589.0 & 66.7 & 11.7 & 4.2 & 0.7 & 17.5 \\
  814:1 & 1590.8 & 35.3 & 12.7 & 2.2 & 0.8 & 35.8 \\
  555:2 & 1470.4 & 16.5 &  7.4 & 1.1 & 0.5 & 44.8 \\
  606:2 & 1467.9 & 14.6 &  9.4 & 1.0 & 0.6 & 63.8 \\
  814:2 & 1468.0 & 19.8 &  7.4 & 1.3 & 0.5 & 37.5 \\
  555:3 & 1486.4 &  6.9 &  6.3 & 0.5 & 0.4 & 90.1 \\
  606:3 & 1493.7 & 12.0 &  6.8 & 0.8 & 0.5 & 57.0 \\
  814:3 & 1525.2 & 16.3 &  5.5 & 1.1 & 0.4 & 33.3 \\
  555:4 & 1488.6 & 14.2 &  6.9 & 1.0 & 0.5 & 48.2 \\
  606:4 & 1493.2 & 15.6 &  6.4 & 1.0 & 0.4 & 40.9 \\
  814:4 & 1507.0 & 28.9 &  5.8 & 1.9 & 0.4 & 20.2 \\
\hline  
\end{tabular}  
\label{tab:rms_corr}
\end{table}

The third order coefficients from Tab. \ref{tab:third_order_terms}, the higher-order distortion maps and a Fortran subroutine to 
apply these corrections are included as a digital archive accompanying this manuscript.

\section{Summary and Future Work}\label{sec:summary}

We explore systematic distortions in WFPC2 exposures, in filters F555W, F606W and F814W. These are beyond the previously determined 
34th-row correction and nominal cubic distortion characterized by AK99 and AK03. 
Our study examines high- and low-frequency spatial systematics by stacking hundreds of 
thousands of residuals from transformations
into three standard catalogs, and from relative transformations between WFPC2 exposures.
In the high-frequency domain, to a resolution of 10 pixels, we find no other systematics beyond the well-known 34th-row correction.
In the low-frequency domain, on scales of 50 pix and larger, we find significant systematics beyond the nominal AK03 distortion.
We characterize these systematics with third-order coefficients and distortion maps presented for each filter and chip.
{\it Gaia } EDR3 is crucial to this calibration, being the one catalog that allows for such an extensive mapping of the geometric
distortions in WFPC2.

The modifications to the nominal (AK03) WFPC2 distortion corrections we present range in size from less than 10 mpix up to
$\sim$50 mpix.
For an individual star, random measuring errors typically will be or order 20 mpix.
The effects of CTE and exposure-to-exposure variation of distortions across a chip can be expected to be of this size or more.
For this reason, the corrections we present are of most value in statistical studies involving many stars and multiple exposures.

As part of the distortion exploration, we perform a comparative analysis of five centering algorithms for WFPC2 star images. 
This allowed us to better understand how each algorithm performs, and to
select the most appropriate algorithm for our purpose. 
A direct comparison of 2DG-derived positions with those from the ePSF algorithm demonstrated spatial variations between the two that
must be accounted for.
The modifications to the nominal AK03 distortion corrections we present are applicable to ePSF-derived positions. 

An interesting finding of our centering-algorithm tests is that PSFEx is a promising astrometric choice, or at 
the least one that can be used in some specific targeted astrometric studies, given its computational needs.  
An issue that remains to be better studied is that of the pixel-phase bias, which operates across the scale of a pixel. 
It is a consequence of detector undersampling and the unavoidable uncertainty in the assumed profile of stellar images, and is
most critical for the WF chips. 
Although the ePSF algorithm performs well in this regard, the pixel-phase effect is still present at some level.
Correction for the effect should be improved upon, if highest astrometric precision is to be achieved.
This will be the topic of a future study. 
The maps built and presented in the current study, however, are not affected by this effect as its contribution is effectively as
random noise participating at a level well below that of other sources of random error.

Finally, an important aspect that remains to be explored in a future study is that of charge transfer efficiency, CTE, 
which can have considerable astrometric impact, especially for short exposures with WFPC2.
While the size of CTE-induced offsets can be considerable, their effect on the purely geometric corrections presented here is
to add a negligible amount of uncertainty due to marginalizing over stellar magnitude.
Our hope is to employ {\it Gaia} EDR3 in the future to derive an empirically based procedure for CTE correction of WFPC2.

\acknowledgments
This work was supported by grant HST-AR-15632 from the Space Telescope Science Institute, STScI, operated by AURA.
We wish to thank the MAST personnel for providing the entire WFPC2 data set to us on physical media; ours was an unusual request that 
was not feasible using the online query tool and the MAST personnel accommodated us graciously.

This study has made use of data from the European Space Agency (ESA) mission {\it Gaia} (\url{https://www.cosmos.esa.int/gaia}),
processed by the Gaia Data Processing and Analysis Consortium (DPAC, \url{https://www.cosmos.esa.int/web/gaia/dpac/consortium}).
Funding for the DPAC has been provided by national institutions, in particular the institutions participating in
the {\it Gaia} Multilateral Agreement.

We are grateful to an anonymous referee whose suggestions helped us greatly improve this manuscript over an earlier version.

%

\vspace{5mm}
\facilities{{\it HST} (WFPC2), MAST, {\it Gaia}}

\end{document}